\documentclass[apj]{emulateapj}
\usepackage{multirow}
\usepackage{longtable}
\usepackage{rotating}

\def\gtorder{\mathrel{\raise.3ex\hbox{$>$}\mkern-14mu
             \lower0.6ex\hbox{$\sim$}}}
\def\ltorder{\mathrel{\raise.3ex\hbox{$<$}\mkern-14mu
             \lower0.6ex\hbox{$\sim$}}}

\slugcomment{Draft of \today}

\shorttitle{Fast Rotator Asteroid - A Case For Cohesion}

\shortauthors{Polishook et al.}

\begin{document}

\title{A 2km-size asteroid challenging the rubble-pile spin barrier - a case for cohesion}

\author{D.~Polishook\altaffilmark{1},
N. Moskovitz \altaffilmark{2},
R.~P. Binzel \altaffilmark{3},
B. Burt \altaffilmark{3},
F.~E. DeMeo \altaffilmark{3},
M.~L. Hinkle \altaffilmark{2},
M. Lockhart \altaffilmark{4},
M. Mommert\altaffilmark{5},
M. Person \altaffilmark{3},
A. Thirouin \altaffilmark{2},
C.~A. Thomas \altaffilmark{6,7,8},
D. Trilling \altaffilmark{5,9,10},
M. Willman \altaffilmark{11},
O. Aharonson \altaffilmark{1}
}

\altaffiltext{1}{Department of Earth and Planetary Sciences, Weizmann Institute of Science, Rehovot 0076100, Israel}
\altaffiltext{2}{Lowell Observatory, 1400 West Mars Hill Road, Flagstaff, AZ 86001, USA}
\altaffiltext{3}{Department of Earth, Atmospheric, and Planetary Sciences, Massachusetts Institute of Technology, Cambridge, MA 02139, USA}
\altaffiltext{4}{PO Box 391274, Cambridge, MA 02139, USA}
\altaffiltext{5}{Northern Arizona University, Department of Physics and Astronomy, PO Box 6010, Flagstaff, AZ 86011, USA}
\altaffiltext{6}{NASA Goddard Space Flight Center, 8800 Greenbelt Rd., Greenbelt, MD 20771, USA}
\altaffiltext{7}{NASA Postdoctoral Program, Oak Ridge Associated Universities, PO Box 117, MS 36, Oak Ridge, TN 37831, USA}
\altaffiltext{8}{Planetary Science Institute, 1700 East Fort Lowell, Suite 106, Tucson, AZ 85719, USA}
\altaffiltext{9}{South African Astronomical Observatory, SA}
\altaffiltext{10}{University of the Western Cape, SA}
\altaffiltext{11}{U. Hawaii/Institute for Astrophysics, Honolulu, HI 96822, USA}

\begin{abstract}
	The rubble pile spin barrier is an upper limit on the rotation rate of asteroids larger than $\sim200-300~m$. Among thousands of asteroids with diameters larger than $\sim300~m$, only a handful of asteroids are known to rotate faster than $2.0~h$, all are in the sub-km range ($\le0.6~km$). Here we present photometric measurements suggesting that (60716) {\it 2000 GD65}, an S-complex, inner-main belt asteroid with a relatively large diameter of $2.3^{+0.6}_{-0.7}~km$, completes one rotation in $1.9529\pm0.0002~h$. Its unique diameter and rotation period allow us to examine scenarios about asteroid internal structure and evolution: a rubble pile bound only by gravity; a rubble-pile with strong cohesion; a monolithic structure; an asteroid experiencing mass shedding; an asteroid experiencing YORP spin-up/down; and an asteroid with a unique octahedron shape results with a four-peak lightcurve and a $3.9~h$ period. We find that the most likely scenario includes a lunar-like cohesion that can prevent (60716) {\it 2000 GD65} from disrupting without requiring a monolithic structure or a unique shape. Due to the uniqueness of (60716) {\it 2000 GD65}, we suggest that most asteroids typically have smaller cohesion than that of lunar regolith.


\end{abstract}


\section{Introduction}
\label{sec:Introduction}

\begin{figure}
\centerline{\includegraphics[width=17cm]{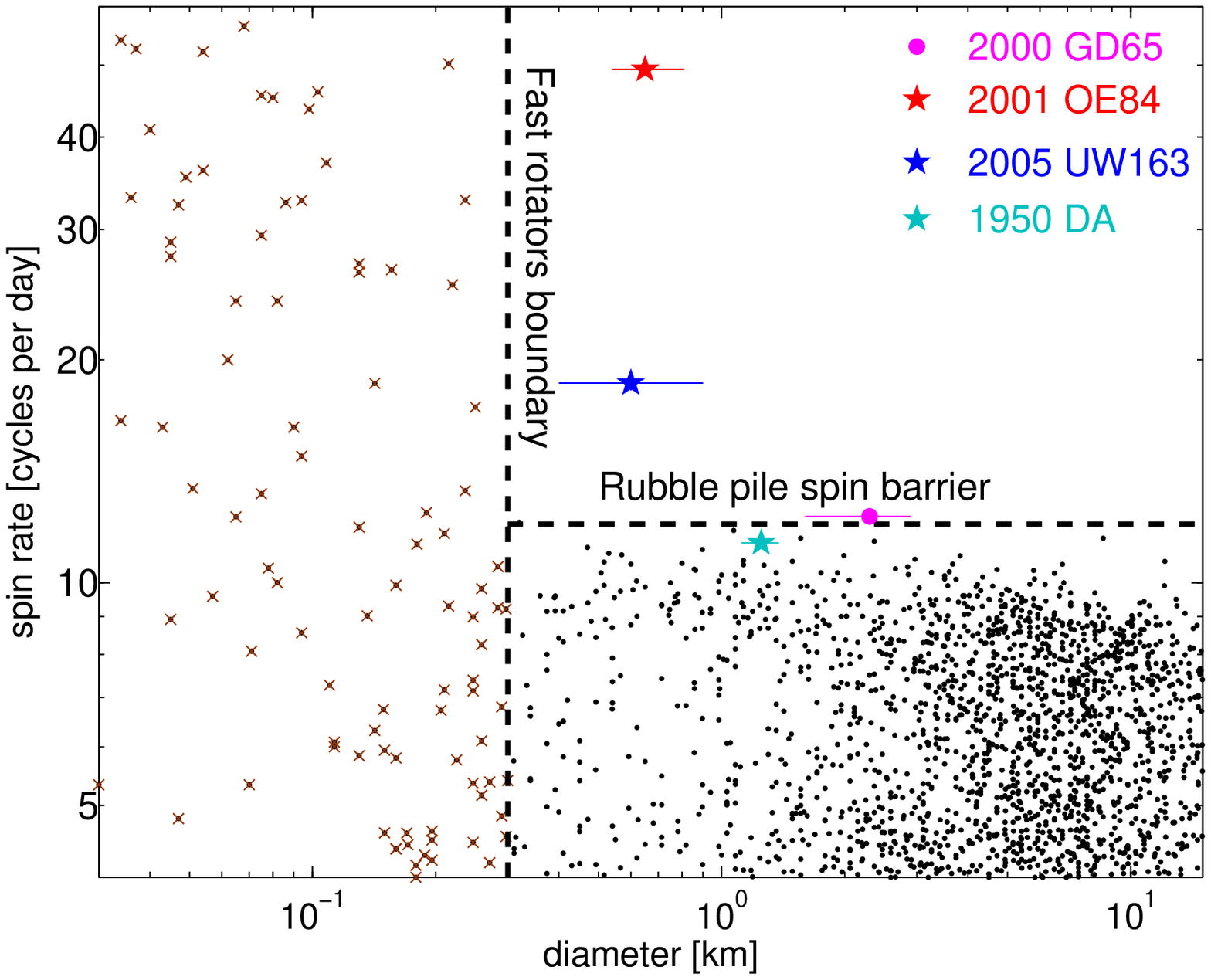}}
\caption{Asteroid diameters vs. spin rates. Asteroids with diameters larger than $300~m$ (black dots) do not rotate faster than $\sim2~h$ while asteroids smaller than $300~m$ (brown crosses) can rotate much faster. The asteroids discussed in the paper are marked: {\it 2000 GD65} (magenta circle) and {\it 1950 DA} (green star) locate on the spin barrier while {\it 2005 UW163} (blue star) and {\it 2001 OE84} (red star) are well within the ``forbidden zone''. The diameters and spin rates are from Warner et al. (2009; database updated for March 2015: {\tiny{http://www.minorplanet.info/datazips/LCLIST\_PUB\_2015MAY.zip}}). Only asteroids with spins of high quality (2 and 3) are displayed. The uncertainty on the diameter is plotted for the four marked asteroids while for the background asteroids the uncertainty is omitted for display reasons and it is estimated to be $\sim30~\%$ of the diameter value. The uncertainty on the rotation periods are smaller than the marker size. The values for the four marked asteroids are from Pravec et al. 2002b ({\it 2001 OE84}), Chang et al. 2014 ({\it 2005 UW163}), Rozitis et al. 2014 ({\it 1950 DA}) and this work ({\it 2000 GD65}).
\label{fig:DiamSpinDiag}}
\end{figure}

Plotting asteroid diameters against their spin rates (Fig.~\ref{fig:DiamSpinDiag}), a clear boundary is noticeable - asteroids larger\footnote{Previous studies put the boundary at $\sim150~m$ (Pravec \& Harris 2000), or $170~m$ (Statler et al. 2013) but new data (Warner's lightcurve database, version of March 2015: {\tiny{http://www.minorplanet.info/datazips/LCLIST\_PUB\_2015MAY.zip}}) add five fast rotators between $200~m$ to $300~m$. Even though there is a large uncertainty on the diameter of these asteroids, we set the size limit conservatively at $300~m$, since it is not crucial for the asteroid we describe here, with a diameter of about $2~km$.}  than $300~m$ do not rotate faster\footnote{In the literature the boundary is estimated to be located between $2$ (Pravec et al. 2002a) and $2.2~h$ (Pravec \& Harris 2000). There are ten more asteroids with rotation periods ranging between $2$ and $2.2~h$. However, estimating their density using Eq. 1 demonstrate they might be rubble piles with bulk density lower than the average density of ordinary chondrites. For asteroids with diameters larger than $\sim10~km$ the spin barrier reduces to slower rotation rates.} than $2~h$  (Pravec et al. 2002a). First noticed by Harris (1996) from a database of 688 asteroids, it was suggested that asteroids have low tensile strength that cannot remain bound against a centrifugal acceleration associated with a rotation period smaller than $\sim2~h$. Subsequent studies on periods of thousands of asteroids (e.g., Pravec \& Harris 2000, Warner et al. 2009, Waszczak et al. 2015) show a similar behavior for asteroids with well-established ($U\ge2$)\footnote{Warner et al. (2009) divide the quality of lightcurves (U) into 3 categories, when U=1 means a period that might be completely wrong, U=2 means the spin is based on less than full coverage, hence it may be wrong by an integer ratio, and U=3 denotes a secure result.} rotation periods. This demonstrates that asteroids larger than $\sim300~m$ are collections of rocks, boulders and dust loosely consolidated by gravity alone and are therefore often referred to as ``rubble piles'' (Chapman 1978). The fact that asteroids smaller than $\sim300~m$ can rotate much faster\footnote{Up to a record of seconds; {\it 2014 RC} has the record with $15~seconds~per~cycle$; NASA/JPL NEO Program Office: {\tiny{http://neo.jpl.nasa.gov/news/news185.html}}.} suggests they are monolithic in nature and might constitute the blocks rubble piles are made of. Alternatively, these small-sized asteroids with extremely fast rotation might be ``rubble piles'' as well, held together by strong cohesion controlled by van der Waals forces and friction between constituent regolith grains (e.g., Holsapple 2007, Goldreich \& Sari 2009, Scheeres et al. 2010, S{\'a}nchez \& Scheeres 2014).

Since sunlight can efficiently apply a thermal torque on asteroids (the YORP effect; Rubincam 2000, Vokrouhlick{\'y} et al. 2015) with diameters smaller than $10~km$ (Pravec \& Harris 2000, Polishook \& Brosch 2009, Jacobson et al. 2014) and spin them up (Rozitis et al. 2013), we understand that asteroids can cross the spin barrier. Excluding less than a handful of cases described below, we do not observe asteroids spinning faster than this barrier, thus we assume that most asteroids with diameters ranging from $\sim300~m$ to $\sim10~km$ are rubble piles. Since most binary asteroids (Pravec et al. 2006) and separated asteroid pairs (Pravec et al. 2010) are located near the rubble pile spin barrier it can be deduced that this is a real boundary and that rubble pile asteroids break apart when the YORP effect accelerates them beyond it. Alternatively, these asteroids can reduce their spin rate by shedding mass and conserving angular momentum.

Presently only two asteroids among thousands with known rotation period and diameter larger than $\sim300~m$ are located beyond the spin barrier\footnote{Conservatively, we excluded the asteroid (144898) {\it 2004 VD17} with $D\sim320~m$ and $P=1.99~h$ (De Luise et al. 2007) since the uncertainty of its parameters might ``put" it in the ``safe zone".}: {\it 2001 OE84} completes a rotation in $0.4865~h$ (Pravec et al. 2002b) and (335433) {\it 2005 UW163} in $1.290~h$ (Chang et al. 2014). Both asteroids are in the sub-km regime ($0.65$ and $0.6~km$, respectively). Recently, Chang et al. (2015) published five additional super-fast rotating candidates (rotation periods ranging between $1.22$ to $1.87~h$) with diameters larger than $\sim300~m$ that were found on the wide-field Palomar Transient Factory (PTF) survey. Even though their fast rotation periods need to be confirmed, and the large uncertainty of their sizes should be reduced\footnote{Since the albedo and taxonomy of these fast rotating candidates are unknown, the uncertainty of their size is $\sim1~km$.}, we should consider if these asteroids form a new type of group. The asteroid reported in this paper, (60716) {\it 2000 GD65} (hereafter {\it GD65}), is the largest in this group, and as such it serves as an extreme case for asteroids that should have failed under their fast spins. Below we present its observations and spin analysis and a handful of models in order to explain its unique parameters.

\section{Observations and measurements}
\label{sec:observations}

\begin{figure}
\centerline{\includegraphics[width=17cm]{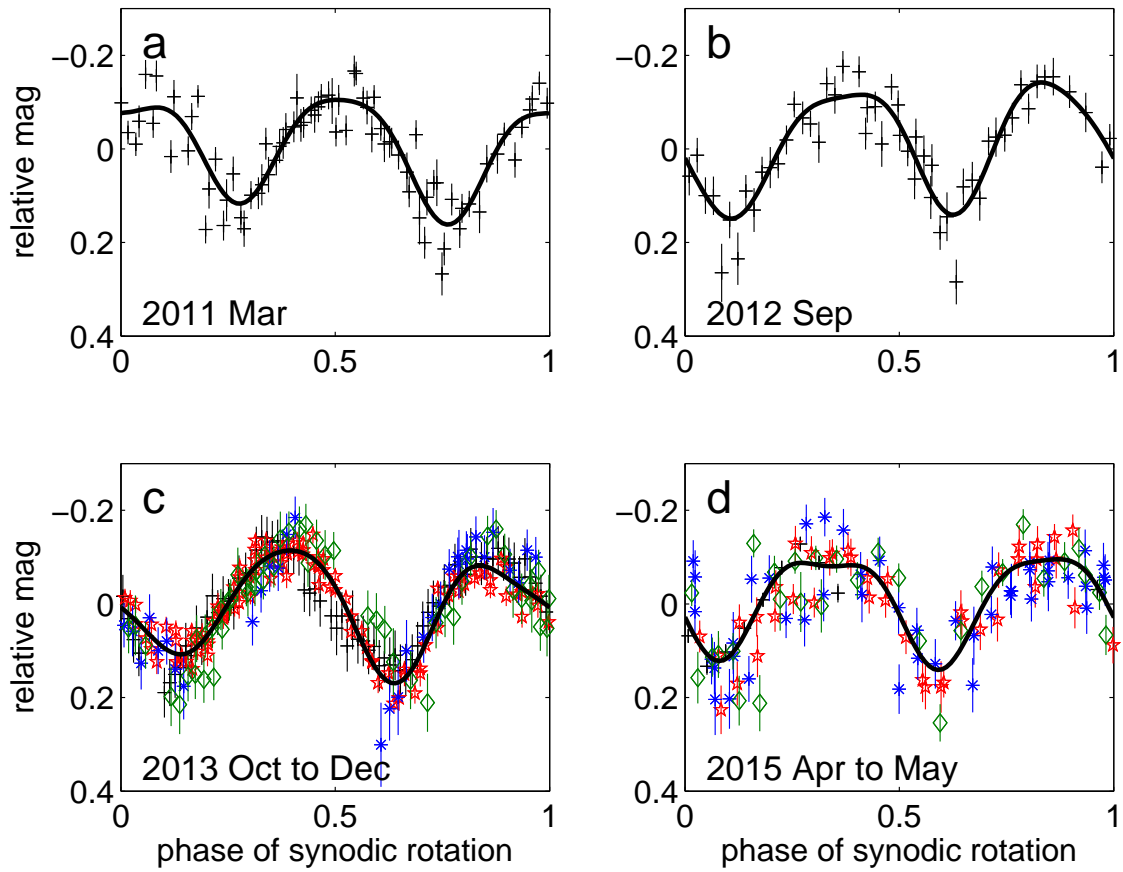}}
\caption{Folded lightcurves of (60716) {\it 2000 GD65} in four apparitions: 2011, 2012, 2013 and 2015 (panels a through d, respectively). See Table I for the observational circumstances and the legend of the markers. Each lightcurve is folded by 1.9529 hours. Amplitude values of the fitted models are 0.27, 0.29, 0.28, 0.24 mag respectively.
\label{fig:Lightcurve}}
\end{figure}

On March 3rd, 2011, {\it GD65} passed next to another target (6070 Rheinland, photometry published in Vokrouhlick{\'y} et al. 2011) within the large field of view (40'x27') of the $0.46~m$ telescope of the Wise Observatory (097, Israel; Brosch et al. 2008). Diagnosed as an extremely fast-rotator (Fig.~\ref{fig:Lightcurve}), we continued observing {\it GD65} on following apparitions with multiple telescopes: Wise's $1~m$ on September 2012, Kitt-Peak's $4~m$ and $2.1~m$ on October 2013, CTIO's $1.3~m$ on December 2013, and Wise's $0.71~m$ and $0.46~m$ on April 2015. For details, see the observing circumstances in Table~\ref{tab:ObsCircum}. Observing time on Kitt-Peak's and CTIO's telescopes was allocated through the Mission Accessible Near-Earth Objects Survey (MANOS; Moskovitz et al. 2014). Reduction included the standard procedures of bias and flat field correction, aperture photometry with 4-pixels radii, and calibration using tens up to hundreds of local comparison stars with a tolerance of $0.02~mag$. The photometry was corrected by light-travel time. For further details on reduction, measurements and calibration refer to Polishook \& Brosch (2009).

\begin{figure}
\centerline{\includegraphics[width=17cm]{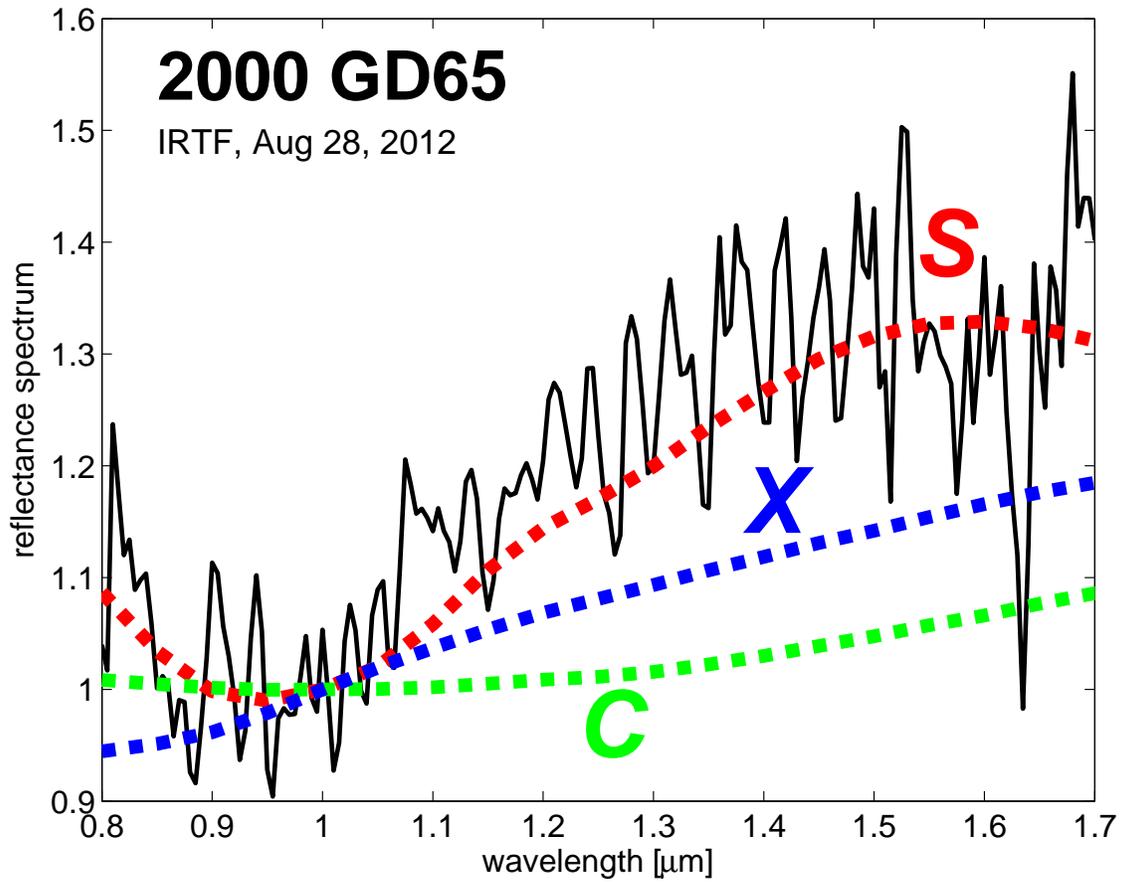}}
\caption{Reflectance spectrum of (60716) {\it 2000 GD65} taken by the IRTF's Spex camera on August 28, 2012. Compared to the three archetypes of asteroids, S- (silicate / ordinary chondrite), C- (carbonaceous ) and X- (metal and other mineralogy) complexes, {\it GD65}'s spectra matches to the S-complex and not the others. All spectra are normalized to unity at 1 $\mu m$. From this measurement we conclude an albedo of $0.197\pm0.051$ (Pravec et al. 2012) and a composition of ordinary chondrites.
\label{fig:Spectrum}}
\end{figure}

In addition to the photometry, near-IR spectroscopy was performed on August 28, 2012 using SpeX, a $0.8-5.4~\mu m$ imager and spectrograph mounted on the NASA Infrared Telescope Facility's (IRTF) $3~m$ telescope (Rayner et al. 2003). Telescope tracking in the spectrograph 0.8'' slit was maintained using the MIT Optical Rapid Imaging System (MORIS; Gulbis et al. 2010) mounted on a side-facing exit window of SpeX. Observing parameters and reduction of the raw SpeX images follow the procedures outlined in DeMeo et al. (2009) and Binzel et al. (2010). This includes flat field correction, sky subtraction, manual aperture selection, background and trace determination, removal of outliers, and a wavelength calibration using arc images. A telluric correction routine was used to model and remove telluric lines. The spectrum was divided by the standard solar analog SA110-361 from Landolt Equatorial Standards list (Landolt 1992) to derive the relative reflectance of the asteroid. Since {\it GD65} was faint during the IRTF observation (Vmag of 19.0) we only present the spectrum from $0.8$ to $1.7~\mu m$. Though noisy, the spectrum best matches an Sq-type taxonomy, which is a sub-type of the S-complex taxonomy (Taxonomy defined by DeMeo et al. 2009; Fig.~\ref{fig:Spectrum}), suggesting it has a composition analogous to an ordinary chondrite{\footnote{Other meteoritic analogs to the S-complex taxonomy of asteroids are rare ($\sim1\%$). They include achondritic meteorites such as acapulcoites, angrites, lodranites, ureilites and winonaites (Burbine et al. 2002), all with grain density of $3.4$ to $3.7~gr~cm^{-3}$. (Macke et al. 2011).}}, although mineralogical modeling is needed for a more detailed compositional analysis.

\section{Spin Analysis}
\label{sec:analysis}

\begin{figure}
\centerline{\includegraphics[width=17cm]{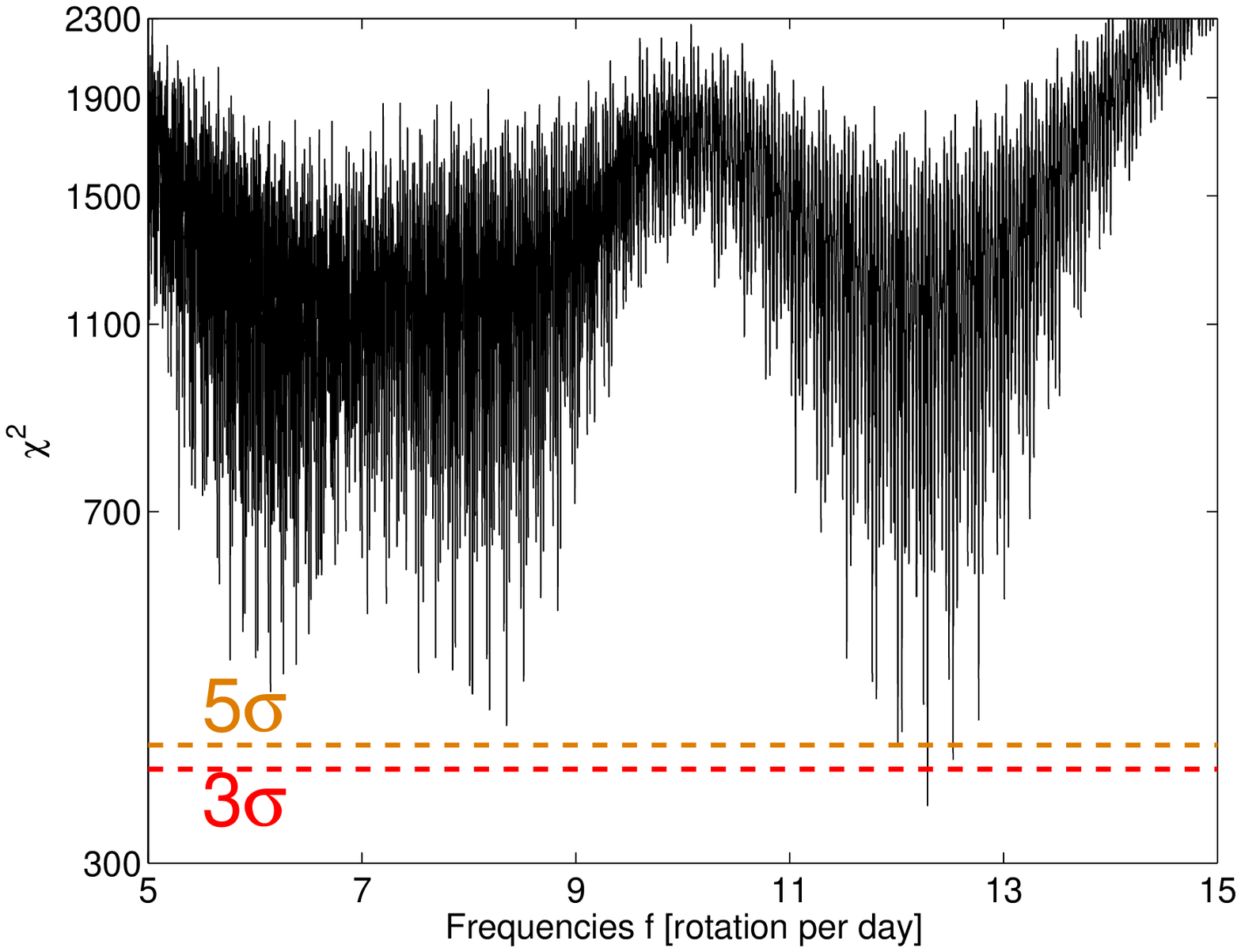}}
\caption{A logarithmic display of the $\chi^2$ values of a range of frequencies tested on the dataset of the 2013 apparition (see Table I for observation circumstances). The red dash-line represents the $\chi^2$ value of $3\sigma$ above the minimal $\chi^2$, and the orange dash-line represents the $5\sigma$. The minimal $\chi^2$ is found for frequency of 12.2894 cycles per day, i.e. a rotation period of 1.9529 hours. Folding the photometric dataset to a lightcurve with 4 peaks (frequency of 6.1447 cycles per day) gives $\chi^2$ values that are significantly higher than the $3$ and $5\sigma$ threshold. A larger frequency of 12.5287 cycles per day (even faster rate) is within the $5\sigma$ uncertainty range (and above the $3\sigma$), but it is rejected since it does not fit the data from the other apparitions.
\label{fig:Chi2}}
\end{figure}

We used a Fourier series analysis to match a wide set of spin frequencies to the lightcurves of {\it GD65} (Polishook et al., 2012). Sets of lightcurves from different apparitions were tested separately. Data from three partly cloudy nights with high systematic error were excluded from the spin analysis. For a given frequency a least-squares minimization was used to derive a $\chi^2$ value. In the $\chi^2$ plane, we tested all local minima that are smaller than the lowest $\chi^2$ + $\Delta\chi^2$, where $\Delta\chi^2$ is calculated from the inverse $\chi^2$ distribution at $3\sigma$ assuming a frequency with 4 harmonics and 2 peaks (see reasoning below). All the matching frequencies from a single apparition were checked on all the data from the other apparitions. This resulted with a single frequency that matches all the measurements and represents a rotation period of $1.9529\pm0.0002~h$. We display the lightcurves from the four apparitions folded by this rotation period (Fig.~\ref{fig:Lightcurve}) and the $\chi^2$ values as a function of different frequencies for the best dataset from the apparition of 2013 (Fig.~\ref{fig:Chi2}). Notably, the lightcurve did not change significantly during the four years of observations even though {\it GD65} was observed in different ecliptic longitude. Since the inclination of {\it GD65}'s orbit is low ($\sim3.2^o$) a lightcurve unchanged during the years suggests an aligned spin axis. We used a code based on the lightcurve inversion technique ({\v D}urech et al. 2010) to derive the spin axis and shape of {\it GD65}, but we did not derive a solution for the spin axis vector that is statistically significant.

Although most known asteroid lightcurves have a 2-peak morphology, some present more complex shapes that are characterized by a larger number of harmonics. Harris et al. (2014) gave examples for such unique lightcurves (e.g., 5404 Uemura with a 6-peak lightcurve) and showed that ``lightcurves with amplitude less than $0.2-0.3~mag$ can be dominated by other harmonics, especially the 4th and 6th.'' This allows us to reject all solutions that are multiples of $1.9529$ excluding the double period ($P=3.9058~h$, 4-peak lightcurve), since the amplitude of the models fitted to {\it GD65}'s lightcurve range between 0.24 to 0.29 magnitude (see Table 1 at Harris et al. 2014). However, is it possible to choose between the 2-peak and the 4-peak solutions?

\begin{figure}
\centerline{\includegraphics[width=12cm]{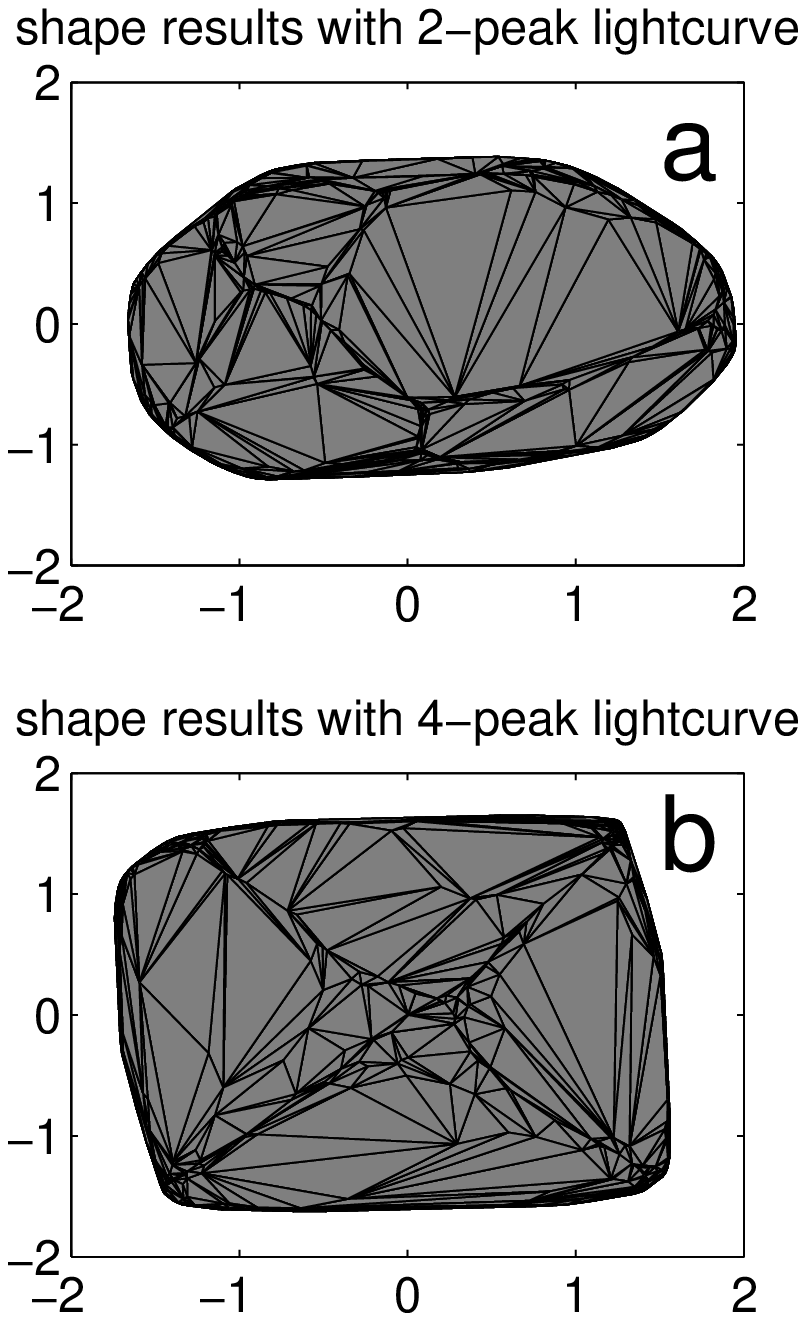}}
\caption{Examples of shape models (top view) derived when fixing the rotation period to 1.9529 hours (a 2-peak lightcurve; frame a) and to 3.9058 hours (a 4-peak lightcurve; frame b). While the shape model of {\it GD65} based on 2-peak lightcurve has a standard looking shape, the shape model based on 4-peak lightcurve have a semi-octahedron shape, with {\bf $\sim$90$^o$ between the sides at the equator}. While the inversion technique results with a convex-hull model for a possible non-convex shape, the ``corners" of the model in frame b are still valid since they are part of the hull of the shape model. This artificial looking shape seems unlikely within our knowledge of asteroid shapes ({\v D}urech et al. 2015), what argues for a 2-peak lightcurve scenario.
\label{fig:RectanguleShape}}
\end{figure}

While the maximal amplitude of {\it GD65}'s lightcurve\footnote{The amplitude of the models fitted to {\it GD65}'s lightcurves at the four apparitions are 0.27 (2011), 0.29 (2012), 0.28 (2013) and 0.24 (2015) magnitude. We decided to use the value of 2011 as the maximal amplitude, since at this apparition the asteroid was observed at phase angle of $2.4^o$, almost at opposition, where the light curve's amplitude is minimal (Zappala et al. 1990).}, $0.27~mag$, is within the mathematical limit of cubical shape ($0.2-0.3~mag$ as determined by Harris et al. 2014), the question arises if such a shape is likely. We used the lightcurve inversion technique again ({\v D}urech et al. 2010), this time with a fixed rotation period of $P=3.9058~h$ (4-peak lightcurve). While we could not derive a single significant solution for the spin axis vector, all the shape models we did derive have a semi-octahedron shape, with {\bf $\sim$90$^o$ between the sides at the equator} (Fig.~\ref{fig:RectanguleShape}, frame b). Not only that such a shape was not seen before on asteroids visited by spacecrafts or resolved by radar observations, this shape seems artificial compared to a standard looking shape model (Fig.~\ref{fig:RectanguleShape}, frame a) that is based on a standard 2-peak lightcurve. While the 4-peak lightcurve solution is valid mathematically, its artificial semi-octahedron shape seems unlikely within our knowledge of asteroid shapes ({\v D}urech et al. 2015).

\begin{figure}
\centerline{\includegraphics[width=17cm]{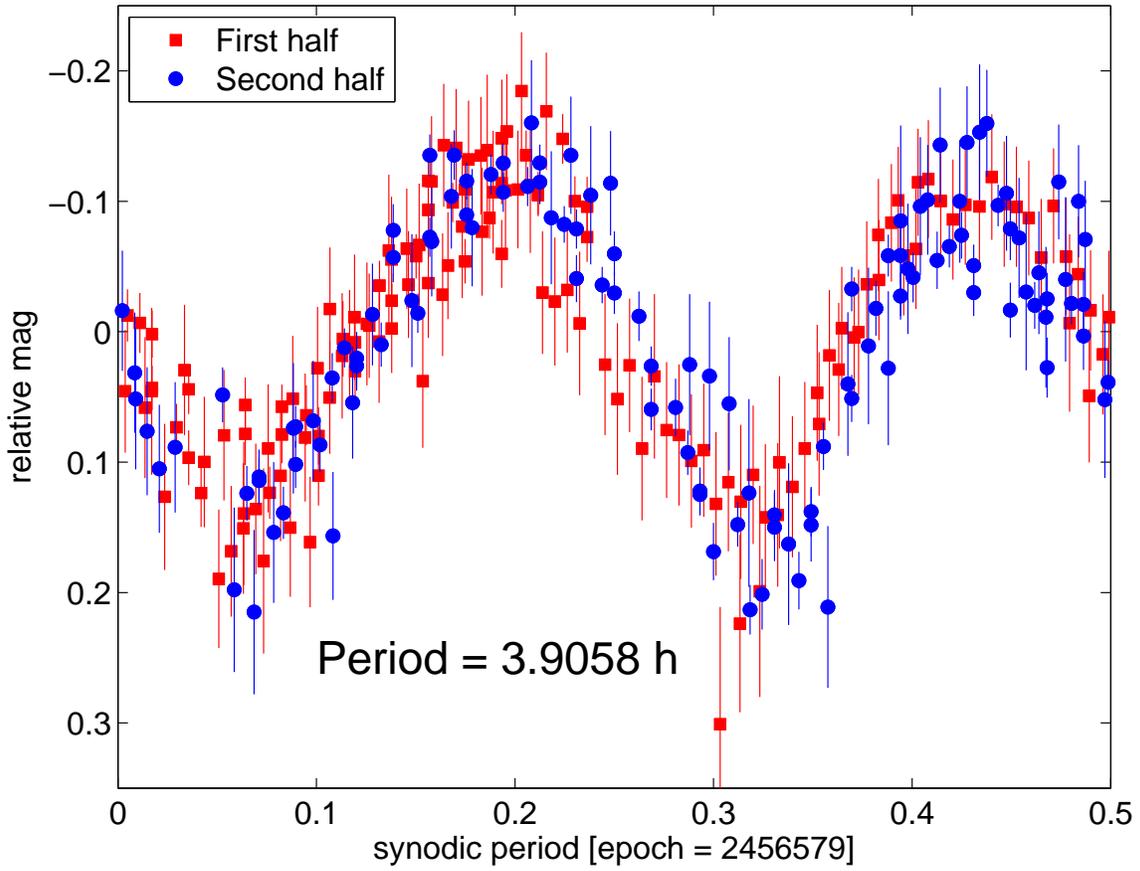}}
\caption{The photometric data from 2013 (the set with smaller magnitude errors) folded by a period of 3.9058 hours (a 4-peak lightcurve) is plotted with the two halves of the rotation phase one on top the other. The two halves present similar lightcurves within the scatter of the data, thus there is no preference of the 4-peak solution over the 2-peak solution.
\label{fig:HalfFold}}
\end{figure}

An additional visual test to confirm a 4-peak lightcurve is to search for morphological differences between different peaks. We folded the 2013 data (the set with smaller magnitude errors) by a period of $3.9058~h$ to construct a 4-peak lightcurve, and plotted the two halves of the rotation phase one on top the other (following Harris et al. 2014; Fig.~\ref{fig:HalfFold}). The two halves of the rotation phase are similar within the scatter of the data, therefore, from a morphological point of view, we cannot prefer the 4-peak over the 2-peak solution.

Finally, we fitted the data of each apparition to a $1.9529~h$ period (2-peak solution) and a doubled period of $3.9058~h$ (4-peak solution) and compared the $\chi^2$ values (Table~\ref{tab:ApparitionChi2}). The fitting was done using 4, 6 and 8 harmonics since a 4-peak lightcurve is defined by the forth harmonic (Harris et al. 2014) and this adds complexity to the fitted model. The data from 2013 and 2015, where {\it GD65} was observed at multiple nights, fits significantly better ($3\sigma$) to the 2-peak solution than the 4-peak solution when 4 and 6 harmonics are used, while fitting the 2011 and 2012 data (a single night per apparition) results with similar $\chi^2$ values (within uncertainty of $3\sigma$). When the eighth harmonics is used, similar $\chi^2$ values are derived for both the 2- and 4-peak scenarios, excluding the 2011 data where the 4-peak solution has a lower $\chi^2$. However, fitting with the eighth harmonic results with over fitting of the model to the data so the systematic errors dominate the model, which is obviously wrong. Therefore, comparing the $\chi^2$ of different fits results with an ambiguous result that cannot prefer the 4-peak over the 2-peak solution and vice versa.
	
Since we cannot confirm the double period of a 4-peak lightcurve from a morphological and a least-squares minimization points of view, we are left with the unlikely shape model constrained by a lightcurve with four peaks and amplitude circa to the mathematical limit of a cube. Therefore, we conclude that the 2-peak solution is the most probable one, even though additional observations should be conducted to reject the scenario of a 4-peak lightcurve.

\section{Discussion}
\label{sec:discussion}

We use the fast rotation of {\it GD65}, relative to its size, in order to constrain its interior structure and its evolution state, by testing 5 scenarios:

i.)	A rubble pile bound only by gravity.

ii.)	A rubble-pile with strong cohesion.

iii.)	A monolithic structure.

iv.)	An asteroid experiencing mass shedding.

v.)	An asteroid experiencing YORP spin-up/down.

We discuss the physics of each model and their implications within the context of asteroid properties.

\subsection{A rubble pile bound only by gravity}
\label{sec:rubblePile}

Assuming {\it GD65} has the same rubble pile, strengthless, structure as other asteroids with similar sizes, one can determine a minimal limit for its bulk density ($\rho$) by assuming its spin period is equal or smaller than its critical rotation period ($P_{crit}=2\pi/\omega_{crit}$) at which the gravitational acceleration equals the centrifugal acceleration:
\begin{equation}
\frac{Gm}{r^2} = \omega^2_{crit}r \Rightarrow P^2_{crit} = \frac{3\pi}{G\rho}(\frac{a}{b})
\label{eq:rubblePile}
\end{equation}

where $r$ and $m$ are the radius and mass of the asteroid and $G$ is the gravitational constant. If the body is elongated, the term a/b is introduced in Eq. 1 (Pravec \& Harris 2000) which is the ratio between two of the three physical axes of the asteroid shape {\it a}, {\it b}, and {\it c}, when $a \ge b \ge c$ (and the asteroid rotates around the shortest axis c).

The relevant parameters for {\it GD65} are the rotation period $P$ ($1.9529\pm0.0002~h$) and the triaxial shape ratio $a/b_{min}$ that can be constrained from the average lightcurve amplitude $\Delta M$ ($0.27\pm0.05~mag$) using $a/b_{min} = 10^{0.4\Delta M} = 1.28\pm0.06$. These parameters limit the density to a minimal value of $\rho_{bulk}=3.7\pm0.2~gr/cm^3$ which is larger than the material density of ordinary and carbonaceous chondrite alike ($\sim 3.3$, $\sim 2.6~gr/cm^3$ respectively; Carry 2012). This minimal value is even larger than the average density of measured S-complex asteroids ($2.72\pm0.54~gr/cm^3$; Carry 2012), which in turn, is lower than the density of ordinary chondrite since most of the S-complex asteroids with known density have a rubble pile structure.

Meteorites with relevant large densities are irons or stony-irons ($\rho_{material} \sim 7.3$, $\sim 4.4~gr/cm^3$ respectively; Carry 2012). However, the S-complex spectrum of {\it GD65} does not match an iron composition (classified as M-type) but rather that of ordinary chondrite with lower density ($\sim 3.3~gr/cm^3$). Therefore, it is likely that {\it GD65} is not a rubble pile bound by the gravity force alone.

An alternative notion might suggest that {\it GD65} is an iron asteroid covered with a siliceous ordinary chondrite surface layer. Indeed, micro-scale compositional heterogeneity was found on some meteorites\footnote{e.g., the Almahata Sitta meteorites (Bischoff et al. 2010) and the meteorites resulted from the Bene{\v s}ov bolide (Spurn{\'y} et al. 2014).} but such a scenario is inconsistent in the macro scale (for asteroids in the $\sim1$ to $\sim100~km$ size range) where the commonly held view is that the composition of an asteroid is approximately homogenous throughout the interior. This is deduced, for example, from the albedo and spectral homogeneity of dynamical families of asteroids, formed by catastrophic collisions (Mothe-Diniz \& Nesvorny 2008, Parker et al. 2008, Masiero et al. 2013) and from the bulk density and composition measurements of {\it Itokawa} by the {\it Hayabusa} spacecraft (Fujiwara et al. 2006).

\subsection{A rubble-pile with a strong cohesion}
\label{sec:cohesion}

Studies of granular physics have demonstrated that van der Waals cohesive forces between small size grains trapped in matrix around larger boulders can enlarge the asteroid internal strength and act as a mechanism resisting rotational disruption (Scheeres et al. 2010, S{\'a}nchez \& Scheeres 2014, Scheeres 2015). Following Holsapple (2004, 2007) and Rozitis et al. (2014) we apply the Drucker-Prager yield criterion that calculates the shear stress in a rotating ellipsoidal body at breakup that constrain the body's cohesion. The average shear stresses in three orthogonal directions ($\sigma_x$, $\sigma_y$, $\sigma_z$), are equal to both gravitational and rotational accelerations:

\begin{equation}
\bar{\sigma}_x = (\rho\omega^2 - 2\pi\rho^2 GA_x)\frac{a^2}{5}
\label{eq:rubblePile}
\end{equation}

\begin{equation}
\bar{\sigma}_y = (\rho\omega^2 - 2\pi\rho^2 GA_y)\frac{b^2}{5}
\label{eq:rubblePile}
\end{equation}

\begin{equation}
\bar{\sigma}_z = (- 2\pi\rho^2 GA_z)\frac{c^2}{5}
\label{eq:rubblePile}
\end{equation}

where $\rho$ is the bulk density, $\omega$ is the spin rate, $a \ge b \ge c$ are the three physical axes of the asteroid and $G$ is the gravitational constant. The terms $A_x$, $A_y$ and $A_z$ are dimensionless functions that depend only on the asteroid triaxial ratios\footnote{From Holsapple (2007):
{\tiny{
\begin{equation}
A_x = \alpha\beta\int_{0}^{\infty}\frac{du}{(u+1)^{3/2}(u+\beta^2)^{1/2}(u+\alpha^2)^{1/2}},
\label{eq:AxAyAz}
\end{equation}
\begin{equation}
A_y = \alpha\beta\int_{0}^{\infty}\frac{du}{(u+1)^{1/2}(u+\beta^2)^{3/2}(u+\alpha^2)^{1/2}},
\label{eq:AxAyAz}
\end{equation}
\begin{equation}
A_z = \alpha\beta\int_{0}^{\infty}\frac{du}{(u+1)^{1/2}(u+\beta^2)^{1/2}(u+\alpha^2)^{3/2}}.
\label{eq:AxAyAz}
\end{equation}
}}
when $\alpha=c/a$ and $\beta=b/a$. Values for {\it GD65} are $A_x=0.60\pm0.01$, $A_y=A_z=0.698\pm0.005$ }. Using these average stresses, the Drucker-Prager failure criterion is (Holsapple 2007):

\begin{equation}
\frac{1}{6}[(\bar{\sigma}_x - \bar{\sigma}_y)^2 + (\bar{\sigma}_y - \bar{\sigma}_z)^2 + (\bar{\sigma}_z - \bar{\sigma}_x)^2] \le [k - s(\bar{\sigma}_x + \bar{\sigma}_y + \bar{\sigma}_z)]^2
\label{eq:rubblePile}
\end{equation}

where $k$ is the internal cohesion and $s$ is a slope constant determined by the angle of friction $\phi$:

\begin{equation}
s = \frac{2sin\phi}{\sqrt{3}(3-sin\phi)}
\label{eq:rubblePile}
\end{equation}

\begin{figure}
\centerline{\includegraphics[width=17cm]{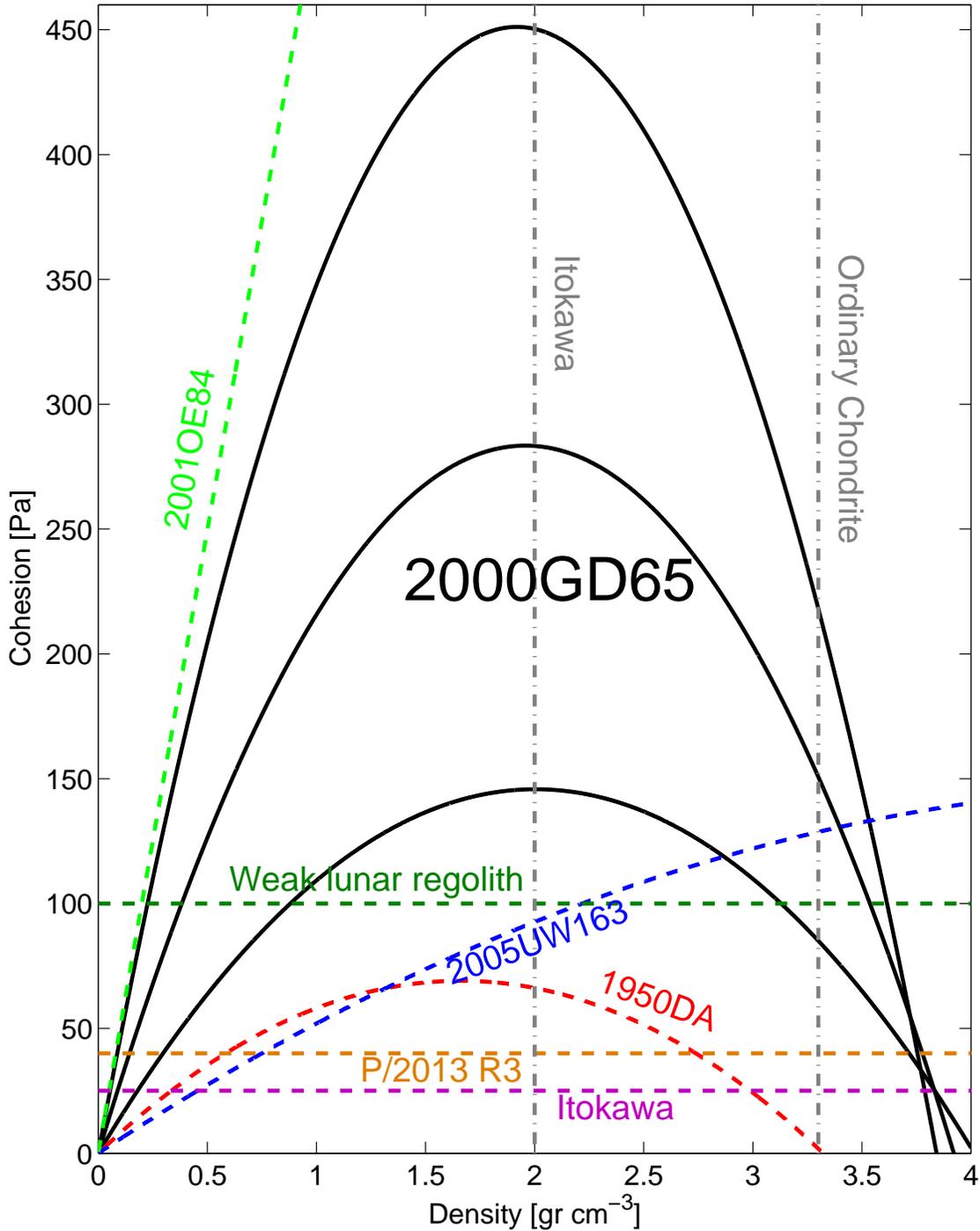}}
\caption{Internal cohesive strength vs. bulk density for the discussed asteroids. The arcs represent the threshold for the cohesion of different asteroids. The mean and range of uncertainty (black solid lines) for {\it 2000 GD65} suggest a cohesion force of $\gtrsim150~Pa$ is needed so {\it GD65} can resist a rotational disruption. This value is just above the minimal values measured on Lunar samples taken by the Apollo missions ($100~Pa$, green dash line, Mitchell et al. 1974), the cohesion estimated for {\it 1950 DA} ($64_{-20}^{+12}~Pa$, red dash line, Rozitis et al. 2014) and {\it 2005 UW163} ($\sim100~Pa$, blue dash line, parameters measured by Chang et al. 2014, while cohesion model is calculated in this work). {\it GD65} cohesion value is also larger than the average cohesion value estimated by S{\'a}nchez \& Scheeres (2014) based on size grains measured on {\it Itokawa} and {\it P/2013 P5} ($\sim25~Pa$, purple dash line) and the lower limit on the cohesion of {\it P/2013 R3}'s progenitor. The cohesion needed to resist rotational disruption of  {\it 2001 OE84} reaches a $\sim700~Pa$ for a rubble pile density of $\sim2~gr~cm^{-3}$ (green dash line), which is still within the range of lunar cohesion ($100-1000~Pa$), therefore, it is not necessary a monolith. The density values of {\t Itokawa}, as an example of a rubble pile asteroid, and the mean density of ordinary chondrite are also marked to limit the possible density range.
\label{fig:Cohesion}}
\end{figure}

We calculate the internal cohesion of {\it GD65} as a function of its bulk density (Fig.~\ref{fig:Cohesion}) using its measured and calculated parameters (Table~\ref{tab:CohesionParam}) and assuming an angle of friction $\phi=40^o$ as measured on lunar regolith (Mitchell et al. 1974). The results show that if {\it GD65} is a standard rubble pile with a density of $\sim2~gr~cm^{-3}$ (like the density of the rubble pile asteroid (25143) {\it Itokawa} measured in situ; Fujiwara et al. 2006), then a cohesive stress of $150$ to $450~Pa$ is needed to keep it from failing (Fig.~\ref{fig:Cohesion}). This supports the notion that asteroids in the $1~km$ size range have internal cohesive strength.

The minimal limit for {\it GD65}'s cohesion is within the measured values of the lunar regolith cohesion, ranging between $100$ to $1000~Pa$ (Mitchell et al. 1974). It is somewhat larger than the cohesion values estimated for specific asteroids (Fig.~\ref{fig:Cohesion}): $64_{-20}^{+12}~Pa$ for (29075) {\it 1950 DA} (Rozitis et al. 2014. Hirabayashi et al. 2015 set this value to range between $75$ to $85~Pa$. Gundlach \& Blum 2015 suggest the value is between $24$ to $88~Pa$), $40-210~Pa$ for the precursor body of the active asteroid {\it P/2013 R3} (Hirabayashi et al. 2014), and about $100~Pa$ for the fast rotator (335433) {\it 2005 UW163} (fast rotation measured by Chang et al. 2014 and cohesion is calculated here using their published parameters). Among km-sized asteroids, the minimal limit for {\it GD65}'s cohesion is only smaller than the minimal cohesion of {\it 2001 OE84} which is about $\sim700~Pa$ for a rubble pile density of $2~gr~cm^{-3}$. This means that {\it 2001 OE84} can be a fast rotating rubble pile with a lunar like cohesion, and it does not necessarily a monolith.

S{\'a}nchez \& Scheeres (2014) indicates that the value of cohesion strength inversely correlates with the grain size in the range of microns. This allowed them to use measurements of grain size of the sampling mission target (25143) {\it Itokawa} (Tsuchiyama et al. 2011) and the active asteroid {\it P/2013 P5} (Jewitt et al. 2013) to derive the average cohesion force among asteroids (in addition to numerical simulations). Since the average grain size for these bodies is $\sim10^{-5}~m$,  S{\'a}nchez \& Scheeres (2014) suggested that the cohesive strength of rubble pile asteroids is $\sim25~Pa$, about an order of magnitude lower than our result.

\begin{figure}
\centerline{\includegraphics[width=17cm]{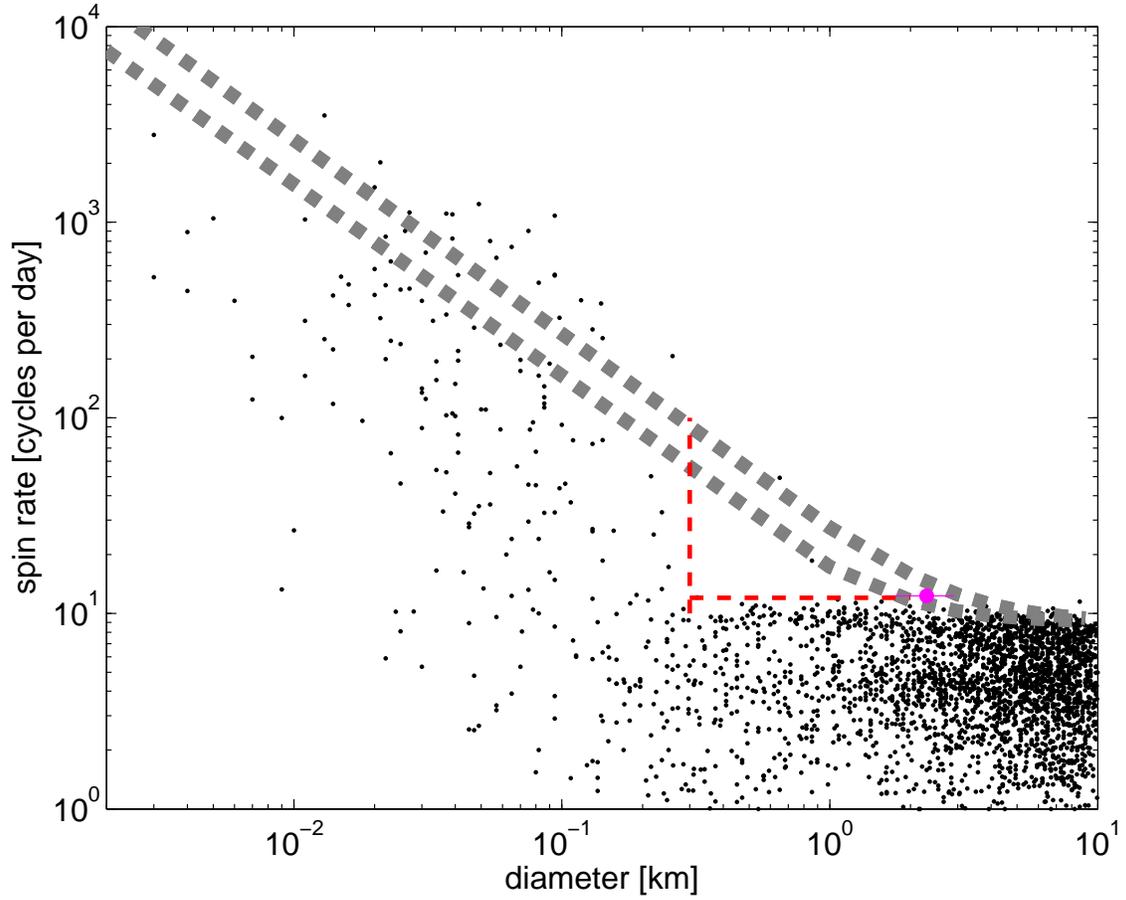}}
\caption{The spin rate as a function of diameter for asteroids with the internal cohesion of {\it GD65} (two grey dash-lines representing the uncertainty in the cohesion value). {\it GD65} is marked (magenta circle), with other asteroids in the background (black points). The deficient in objects in the triangle area below the lines, above a spin period of $2~h$ and right to a diameter of $\sim300~m$ (red dash-lines), rise doubts in the model that all asteroids are rubble piles with internal cohesion, and support the concept that asteroids with $D<\sim300~m$ are monolithic bodies.
\label{fig:CohesionSizeLimit}}
\end{figure}

This range of values can suggest two possible options: 

1. Most (if not all) asteroids in the $1~km$ size range have internal cohesion similar to weak lunar regolith (Mitchell et al. 1974). {\it GD65} and the other mentioned asteroids ({\it 1950 DA}, {\it 2005 UW163} and {\it P/2013 R3}'s precursor) just happened to rotate fast compared to other asteroids, thus revealing the cohesion value. Likewise for {\it Itokawa} and {\it P/2013 P5}, that might have larger cohesion than estimated by their grain size\footnote{{\it Itokawa} and {\it P/2013 P5} might have exhibited larger particles at the surface than the finer particles within (for example, due to the so-called Brazil nut effect; Tancredi et al. 2015), thus misleading S{\'a}nchez \& Scheeres (2014) to underestimate the value of cohesion an order of magnitude.}. If true, this means that the regolith within rubble pile asteroids is similar in cohesion value to the regolith on the Moon and that the particle size of the fine grain in asteroid interiors is approximately $10^{-6}~m$ (and not $10^{-5}~m$ as suggested by S{\'a}nchez \& Scheeres, 2014). However, the main problem with this scenario, where {\it GD65} is an example of similar sized asteroids, is that $>99\%$ of the asteroids in the $1~km$ size bin do not rotate as fast as {\it GD65}, leaving its unique rotation unsolved. Moreover, smaller asteroids ($0.3<D<2.3~km$), that should rotate faster than {\it GD65}, are hardly exist (Fig.~\ref{fig:CohesionSizeLimit}), and their deficient rise doubts in the validity of this model.

2. Alternatively, having a significant cohesion makes {\it GD65} (and the other asteroids mentioned above) unique among $1~km$ size asteroids. A possible reason for a high cohesion might be that {\it GD65} is an older body that suffered from more non-catastrophic collisions, micrometeorite bombardment, thermal cracking fracturing (Delbo et al. 2014), etc. Such a history fractured {\it GD65}'s internal structure, reduced the size of the grains within its interior and on its surface, and thereby increased its cohesion. {\it GD65} does not belong to any known dynamical family (Asteroids Dynamic Site, AstDys-2 website\footnote{{\tiny{http://hamilton.dm.unipi.it/astdys.}}}), which also supports an old age. If {\it GD65}'s cohesion is large due to a small, lunar-like grains, its thermal inertia should be small as well. Indeed, {\it 1950 DA}, with cohesion of $64_{-20}^{+12}~Pa$ was found to have thermal inertia of $24_{-14}^{+20}~J~m^{-2}~K^{-1}~s^{-0.5}$ (Rozitis et al. 2014), similar to that of the moon ($\sim45~J~m^{-2}~K^{-1}~s^{-0.5}$; Wesselink 1948), and one order of magnitude smaller than the thermal inertia of km-sized near-Earth asteroids ($200\pm40~J~m^{-2}~K^{-1}~s^{-0.5}$; Delbo et al. 2007).

\subsection{A monolithic structure}
\label{sec:shedding}

If {\it GD65} has a coherent, monolithic structure it can survive fast rotation without failing. For that, its collisional history should have been mild with minimal number of collisions so its interior did not shatter since its formation (as opposed to the scenario in the previous section). Theoretical models (e.g., Farinella et al. 1998, Bottke et al. 2005, de El{\'i}a \& Brunini 2007, Marzari et al. 2011) estimate that the collisional lifetime of a $D\sim1~km$ main belt asteroid, such as {\it GD65}\footnote{{\it GD65} orbits the Sun in the center main belt of asteroids. Perihelion is $2.2~AU$; Aphelion is $2.7~AU$.}, is $\sim0.5\times10^9~years$. This puts a higher limit on the age of {\it GD65}, suggesting that, if it is indeed a monolith, it was formed in an order of a $10^9$ years, following a catastrophic collision of a larger body. However, {\it GD65} does not belong to any known dynamical family (Asteroids Dynamic Site, AstDys-2 website\footnote{{\tiny{http://hamilton.dm.unipi.it/astdys.}}}).

Furthermore, models of catastrophic collisions (Michel et al. 2004) demonstrated that the progenitors of asteroid families are extensively fractured before being disrupted, i.e., they had a rubble pile structure and not a monolithic one. Since the progenitors of asteroid families are rubble piles, the same should be assumed for asteroids outside of dynamical families. Asphaug et al. (1998) showed that when a non-porous, intact body\footnote{As a target, Asphaug et al. (1998) used the shape model of the near-Earth asteroid (4769) Castalia, with a longest dimension of $1.6~km$, which is similar in size to {\it GD65}.} is impacted by a $16~m$ projectile, it partly fractured but do not disperse, by that forming a rubble pile. We used the size frequency distribution constructed by Bottke et al. (2005) to calculate the probability that a $2.3~km$ {\it GD65} be impacted by a $16~m$ projectile. We used the intrinsic collision probability between main belt asteroids from Bottke et al. (1994), when $P_i = 2.85e-18~km^{-2}~yr^{-1}$,
\begin{equation}
N_{impacts} = P_i N (> r_{projectile}) (r_{GD65} + r_{projectile}) ^ 2,
\label{eq:collisionProbability}
\end{equation}
and derived $N_{impacts}=2.4\cdot10^{-5}~yr^{-1}$, meaning that within a $10^9$ years, {\it GD65} suffered from $\sim10^4$ small impacts that made it a rubble pile, in the case it started its existence as a monolithic object. Therefore, the idea that {\it GD65} has managed to avoid collisions and maintained a monolithic nature, appear unlikely.

\subsection{An asteroid experiencing mass shedding}
\label{sec:shedding}

The uniqueness of {\it GD65} might be due to its evolution rather than its structure. {\it GD65} might have been recently pushed beyond the rubble pile spin barrier by the YORP effect. In this scenario, {\it GD65} is an ordinary rubble pile asteroid, under the dominating control of the centrifugal acceleration and therefore should shed mass. To check this possibility we compared our data with previously observed events of mass shedding\footnote{even though different models describe alternative processes with varying amounts of ejected material (Scheeres 2007, Walsh et al. 2008, Pravec et al. 2010).}. Evolving comas or tails around active asteroids (Jewitt 2012) were previously noticeable in data from $1~m$ class telescopes on sub-km asteroids, with visible magnitude of $\sim19.9$ ({\it P/2010 A2}, Birtwhistle et al. 2010), $20.5-21.0$ ({\it P/2013 P5}, Micheli et al. 2013) or $\sim20.0$ ({\it P/2013 R3}, Hill et al. 2013) even without stacking the images. These features were probably formed by the YORP-rotational fission mechanism and they last for many months (Jewitt et al. 2010, 2013, 2014). We would expect to see similar coma or a tail around {\it GD65} if it were indeed shedding mass.

In search for extended emission around the target, we perform point-spread-function (PSF) subtraction from r-band imaging data taken at KPNO 4m in search for extended emission around the target. The KPNO 4m data set (75 individual frames) provides good signal-to-noise and an imaging resolution of 0.26 arcsec per pixel, which makes these data appropriate for this search. Using IRAF/DAOPHOT routines (Stetson 1987) we produce PSF models for each frame based on field sources and subtract these models from the target image. Those residual images for which the subtraction succeeded and that are not affected by background sources (51 frames) are average-combined in the moving frame of the target to improve the signal-to-noise ratio of potential extended emission. In order to reveal extended emission, we determine the radial brightness distribution centered around the target and fit a $1/\rho$ profile to this distribution, which would be expected in the case of continuous optically thin dust emission. We do not find any clear signs of extended emission around the target.

Based on the radial brightness profile, we derive a $3~\sigma$ upper limit on the dust production rate that conforms with our observations using the ``$A f \rho$'' formalism by A'Hearn et al. (1984) and the method used by Mommert et al. (2014).  We assume a dust particle diameter of $5$~mm, which is in accordance with dust particle diameters observed in other mass-shedding active asteroids (e.g., Jewitt et al. 2010, 2014), a dust bulk density of $3.3~g~cm^{-3}$, and a geometric albedo of $0.197$. For the dust velocity we assume an upper limit that is equal to the escape velocity from the surface of {\it GD65}; assuming an upper-limit on the target bulk density of $3.3~g~cm^{-3}$ and a diameter of $2.3~km$, we find an upper limit on the escape velocity of $1.5~m~s^{-1}$ (which is comparable to dust velocities in other active asteroids, e.g., Jewitt et al. 2010, 2014). Based on these assumptions, we find $3~\sigma$ upper limits on $A~f~\rho \le 1~cm$ and the dust production rate $Q \le 2~kg~s^{-1}$. We find an upper limit (3$\sigma$) on the total amount of dust around the target of ${\le}10^7~kg$, which is one order of magnitude lower than the amount of dust in the mass-shedding event of active asteroid {\it P/2013 R3} ($Q \sim 2\cdot 10^8~kg$; Jewitt et al. 2014).

Alternatively, if the mass shedding occurred in the recent past we probably missed the opportunity to observe dusty features, since all the dust has settled or dispersed before we conducted our observation. However, Jacobson and Scheeres (2011) conducted a numerical model of the rotational fission event, and found that the complete disruption of the system, in which the ejected component left the Hill sphere of the asteroid and the asteroid's spin rate is reduced to a ``safe value'' (forming an {\it asteroid pair}; Vokrouhlicky \& Nesvorny 2008, Pravec et al. 2010), is on the order of tens to hundreds of days (see Table~\ref{tab:ApparitionChi2} in Jacobson \& Scheeres, 2011), smaller than the time it takes the dusty features to disperse according observations of active asteroids (Jewitt et al. 2010, 2013, 2014). This reduces the possibility that we missed {\it GD65}'s mass shedding process because currently it is rotating fast.

\subsection{An asteroid experiencing YORP spin-up/down}
\label{sec:YORP}

Alternatively, {\it GD65} might be in the phase of spinning down after mass shedding (Scheeres 2007) or reshaping (S{\'a}nchez \& Scheeres 2012) as it decreases its angular velocity back to a value below the spin barrier. We looked for evolving spin rate in our four-year apparition dataset, but could not detect any significant difference. This is not surprising since spin rate acceleration ($d\omega/dt$) by the YORP effect was found to be on the order of $10^{-3}$ to $10^{-2}~rad/yr^2$ (Rozitis et al. 2013) while the uncertainty of {\it GD65}'s spin value we found\footnote{With rotation period and its uncertainty are $P\pm dP$ and time range between observations is $dt=4~years$: $\omega_1=2\pi /(P+dP)$, $\omega_2=2\pi /(P-dP) \to (\omega_1- \omega_2)/dt = 1.414~rad~y^{-2}$.} is two to three orders of magnitude larger ($\sim1.5~rad/yr^2$). Therefore, we have no way to know if {\it GD65} is spinning-down (or spinning-up) in the last 4 years. Future observations (e.g., September 2016, December 2017 and May 2019) could potentially rule out or confirm this scenario.

\section{Summary and conclusions}
\label{sec:conclusions}

The main belt asteroid (60716) {\it 2000 GD65}, with relatively large diameter of $2.3~km$ and an S-complex taxonomy, completes a single rotation within $1.9529\pm0.0002~h$, just across the edge of the rubble pile spin barrier. Its unique diameter and rotation period allow us to examine different scenarios about asteroids internal structure. Below we summarize our finds about each scenario:

- We reject a rubble pile structure with no cohesive forces between components since the required density to resist rotational disruption does not fit the density estimated from spectroscopic measurements.

- We find that cohesion of $150$ to $450~Pa$, similar to weak lunar cohesion, can resist rotational disruption of {\it GD65} due to its fast rotation. The cohesion value is 2 to 10 times larger than the value of other asteroids measured for their cohesion (e.g., {\it Itokawa}, {\it 1950 DA}, {\it 2005 UW163}, {\it P/2013 P5} and {\it P/2013 R3}'s precursor). We explain this difference and the uniqueness in {\it GD65}'s spin rate by a possible old age of {\it GD65}, in which extended non-catastrophic collisions or other mechanisms (such as thermal cracking; Delbo et al. 2014) could grind grain sizes to an order of $10^{-6}~m$.

- We can neither confirm nor reject that {\it GD65} has a monolithic structure, though dynamical and collisional models make it unlikely.

- We reject the scenario that {\it GD65} is a rubble pile undergoing disintegration, since no dust plumes, coma or tail were observed within detection limits.

- We can neither confirm nor reject whether {\it GD65} is spinning-down after a putative disruption. Further observations are needed to pinpoint a possible spin change.

	Since we confirm that a cohesion-based model can explain {\it GD65}'s stability against fast rotation, we estimate its implication on the entire population of asteroids of similar size. There are about a thousand asteroids with well-known rotation periods in the size range of {\it GD65}. Including {\it 2001 OE84}, {\it 2005 UW163} and perhaps the precursor of {\it P/2013 R3}, there are only three\footnote{After confirming the fast rotation periods of the additional \newline 5 candidates recently measured by Chang et al. (2015) and accurately measuring their diameters they might fit to this category as well.} asteroids with rotation periods fast enough to allow estimation of a large cohesion higher than $100~Pa$. This is similar to weak-lunar regolith and larger than the values estimated for other asteroids by a factor of 2 to 10. Since the cohesion is correlated with grain size, and since the fraction of such fast rotating asteroids is $<1\%$ among the list of measured asteroids, we conclude that the regolith particles on and within most asteroids are larger than those on the moon by a factor of 2 to 10 as well.
	Alternatively, we should note that some of the other models are not overwhelmingly denied. Future photometric observations should verify that {\it GD65} does not complete a rotation in the double period of $3.9058~h$ and that its spin does not alter. Mineralogical analysis could be performed on its reflectance spectrum with wider wavelength range and higher S/N to better characterize its composition and density. Moreover, the question of how unique is the combination of {\it GD65}'s spin and size should be tested by measuring the exact size, shape and composition of other fast rotating asteroids ($P<2.5~h$) and by conducting large photometric surveys (e.g., Waszczak et al. 2015) in order to see if {\it GD65} represents an unfamiliar group or is anomalous.

\acknowledgments

We are grateful to the referees for their thorough reports that improve the manuscript. We thank William Bottke and Dave O'Brien for their insights and fruitful discussion. DP is grateful to the ministry of Science, Technology and Space of the Israeli government for their Ramon fellowship for post-docs, and the AXA research fund for their generous post-doc fellowship, during the years of observations and analysis. FED acknowledges funding from NASA under grant number NNX12AL26G. CAT was supported by an appointment to the NASA Postdoctoral Program at Goddard Space Flight Center, administrated by Oak Ridge Associated Universities through a contract with NASA. OA would like to acknowledge support from the Helen Kimmel Center for Planetary Science and the ISF I-CORE program "Origins: From the Big Bang to Planets".

We acknowledge support from NASA NEOO grant number NNX14AN82G, awarded to the Mission Accessible Near-Earth Object Survey (MANOS). We are thankful to the Wise Observatory staff for their continuous help and generous time allocation. Observations for this study were performed in Arizona, Chile, Hawaii and Israel. The people of all nations that support hosting professional observatories are praised for understanding and supporting the importance of astronomical studies.

\begin{deluxetable*}{ccccccccc}
\tablecolumns{9}
\tablewidth{0pt}
\tablecaption{Circumstances of observations}
\tablehead{
\colhead{Date} &
\colhead{Telescope} &
\colhead{Filter} &
\colhead{Time span} &
\colhead{Image Num.} &
\colhead{${\it r}$} &
\colhead{${\it \Delta}$} &
\colhead{${\it \alpha}$} &
\colhead{Marker on Fig.~\ref{fig:Lightcurve}} \\
\colhead{}          &
\colhead{}          &
\colhead{}          &
\colhead{[hour]}    &
\colhead{}          &
\colhead{[AU]} &
\colhead{[AU]} &
\colhead{[deg]} &
\colhead{}
}
\startdata
20110303 & Wise 0.46~m & Clear & 6.64 & 77 & 2.23 & 1.24 & 2.4 & panel a, black cross \\
20120828 & IRTF 3~m & Spectroscopy: $0.8-1.6 \mu m$ & 0.70 & 18 & 2.52 & 1.55 & 8.5 & Fig.~\ref{fig:Spectrum}, black line \\
20120914 & Wise 1~m & Clear & 4.21 & 53 & 2.53 & 1.68 & 15 & panel b, black cross \\
20131013 & Kitt Peak 4~m & r' & 1.84 & 75 & 2.63 & 1.88 & 16.8 & panel c, black cross \\
20131027 & Kitt Peak 2.1~m & r' & 2.67 & 110 & 2.62 & 1.74 & 12.2 & panel c, red star \\
20131204 & CTIO 1.3~m & V & 1.99 & 33 & 2.60 & 1.63 & 4.8 & panel c, blue asterisk \\
20131206 & CTIO 1.3~m & V & 2.88 & 54 & 2.60 & 1.64 & 5.7 & panel c, green diamond \\
20150423 & Wise 0.71~m & Luminance & 0.77 & 10 & 2.17 & 1.18 & 6.7 & panel d, black cross \\
20150424 & Wise 0.46~m & Luminance & 3.98 & 41 & 2.17 & 1.18 & 7.3 & panel d, red star \\
20150425 & Wise 0.46~m & Luminance & 4.59 & 47 & 2.17 & 1.19 & 7.7 & panel d, blue asterisk \\
20150513 & Wise 0.71~m & Luminance & 2.73 & 30 & 2.17 & 1.28 & 16.3 & panel d, green diamond
 \enddata
\tablenotetext{}{Columns: date of observation, telescope, filter, hourly time span, number of images, heliocentric and geocentric distances, phase angle, and marker shape and color on Fig.~\ref{fig:Lightcurve}.}
\tablenotetext{}{Data from three partly cloudy nights with high systematic error were excluded from the spin analysis: Sep 15 2012, Dec 3 2013, and May 12 2015. }
\label{tab:ObsCircum}
\end{deluxetable*}

\begin{deluxetable*}{ccccccccccccc}
\tablecolumns{13}
\tablewidth{0pt}
\tablecaption{The minimal $\chi^2$ values of the matched dataset of each apparition to a 2- and 4-peak lightcurves using 4, 6, and 8 harmonics.}
\tablehead{
\colhead{Apparition} &
\colhead{4 harmonics} &
\colhead{} &
\colhead{} &
\colhead{} &
\colhead{6 harmonics} &
\colhead{} &
\colhead{} &
\colhead{} &
\colhead{8 harmonics} &
\colhead{} &
\colhead{} &
\colhead{} \\
\colhead{} &
\colhead{2-peak} &
\colhead{} &
\colhead{4-peak} &
\colhead{} &
\colhead{2-peak} &
\colhead{} &
\colhead{4-peak} &
\colhead{} &
\colhead{2-peak} &
\colhead{} &
\colhead{4-peak} &
\colhead{} \\
\colhead{} &
\colhead{$\chi^2$} &
\colhead{$\Delta\chi^2$} &
\colhead{$\chi^2$} &
\colhead{$\Delta\chi^2$} &
\colhead{$\chi^2$} &
\colhead{$\Delta\chi^2$} &
\colhead{$\chi^2$} &
\colhead{$\Delta\chi^2$} &
\colhead{$\chi^2$} &
\colhead{$\Delta\chi^2$} &
\colhead{$\chi^2$} &
\colhead{$\Delta\chi^2$}
}
\startdata
2011 & 183 & 27 & 173 & 27 & 173 & 33 & 148 & 33 & 166 & 39 & 112 & 39 \\
2012 & 74 & 27 & 84 & 27 & 63 & 33 & 62 & 33 & 61 & 39 & 48 & 39 \\
2013 & 358 & 32 & 461 & 32 & 354 & 38 & 395 & 38 & 349 & 44 & 344 & 44 \\
2015 & 272 & 32 & 325 & 32 & 263 & 38 & 306 & 38 & 256 & 44 & 241 & 44
\enddata
\tablenotetext{}{The $\Delta\chi^2$ is calculated from the inverse $\chi^2$ distribution at $3\sigma$. See text for more details.}
\label{tab:ApparitionChi2}
\end{deluxetable*}

\begin{deluxetable*}{lll}
\tablecolumns{3}
\tablewidth{0pt}
\tablecaption{physical parameters of (60716) {\it 2000 GD65}}
\tablehead{
\colhead{Parameters} &
\colhead{Values} &
\colhead{Reference}
}
\startdata
Absolute Magnitude H & $15.6^{+0.542}_{-0.242}$ & MPC website\footnote{{\tiny{http://www.minorplanetcenter.net/iau/mpc.html}}}. Error is estimated by Pravec et al. (2012). See their Fig. 1.  \\
Spectral classification& S-complex & Our spectral observations \\
Albedo $P_v$ & $0.197\pm0.051$ & Average value for S-complex asteroids (Pravec et al. 2011). \\
                           &                             & {\it GD65} was not observed by WISE/NEOWISE. \\
Diameter D & $2.3^{+0.6}_{-0.7}~km$ & $D=\frac{1329}{sqrt{P_V}}10^{-0.2H}$ \\
Rotation Period P & $1.9529\pm0.0002~h$ & Our photometric observations \\
Mean lightcurve amplitude & $0.27\pm0.05~mag$ & Our photometric observations \\
Triaxial ratios & $a/b=1.28\pm0.06$ & $a/b=10^{0.4\Delta M}$ \\
                         & $b/c=1.0$                   & $b/c=$ minimal value is assumed in order to get the minimal cohesion value \\
Triaxial semi-axes & $a = 1.3^{+300}_{-400}~km$ & $\frac{4\pi}{3}(abc)^3=\frac{4\pi}{3}(\frac{D}{2})^3$ \\
                                  & $b = 1.0\pm0.3~km$ &  \\
                                  & $c = 1.0\pm0.3~km$ &
\enddata
\label{tab:CohesionParam}
\end{deluxetable*}

\end{document}